\documentclass[%
twocolumn,%
showpacs,%
showkeys,%
preprintnumbers,%
amsmath,amssymb,%
page%
]{revtex4}
\usepackage{graphicx}
\usepackage{dcolumn}
\usepackage{bm}

\def\plus{_+}        
\def\minus{_-}       
\def\plusminus{_\pm} 

\def\timeevol#1{\setbox1=\hbox{$\longmapsto$}\setbox2=\hbox{$
    \scriptstyle #1$}\copy1\kern-.5\wd1\kern-.5\wd2
    \raise-1.2\ht2\copy2\kern-.5\wd2\kern\wd1}
\def\d{{\rm d}}  

\begin{document}


\title{Quantum Abacus}
\author{
Taksu Cheon${ }^{1}$,
Izumi Tsutsui${ }^{2}$,
and
Tam\'{a}s F\"{u}l\"{o}p${ }^{2}$
}

\affiliation{
${ }^1$
Laboratory of Physics, Kochi University of Technology,
Tosa Yamada, Kochi 782-8502, Japan\\
${ }^2$
Institute of Particle and Nuclear Studies,
KEK, Tsukuba, Ibaraki 305-0801, Japan
}

\date{April 7, 2004}

\begin{abstract}
We show that the $U(2)$ family
of point interactions on a line can be utilized to provide the $U(2)$
family of qubit operations for quantum information processing.
Qubits are realized as localized states in either side of the point interaction 
which represents a controllable gate.
The manipulation of qubits proceeds in a manner 
analogous to the operation of an abacus.
\end{abstract}

\pacs{03.67.Lx, 03.65.Db, 81.07.Vb}
\keywords{quantum computation, quantum contact interaction, quantum wire}

\maketitle

The idea of representing numbers as spatial locations is probably
as old as mathematics itself.
Even today, a functioning example of locational representation of
numbers is found in the form of a calculator known as
{\it abacus},
in which digits, or bits, are stored as locations of one-dimensionally
mobile objects.
The purpose of this note is to present a quantum version of the abacus,
a model system in which {\it qubits} are realized
as spatially localized states in a one-dimensional space
that is divided into two identical regions by infinitely thin barrier.
The key to this realization is the concept of
quantum point interactions \cite{SE86,AG88,CS99,AK00},
which is a mathematical abstraction of thin barriers acting on quantum particles.
It is shown that the $U(2)$ parameter family of all possible quantum point interactions, 
with the use of time evolution operators, 
can be turned into the $U(2)$ structure of operations on a two
dimensional qubit Hilbert space.
Specifically, the $S^2$ subfamily formed by the scale-invariant point
interactions is turned into the Bloch sphere of qubit transformations.

The standard realization of a qubit Hilbert space employs
a system with two eigenstates, a spin one-half system
being its archetype \cite{NC00}.
Being purely quantal in nature, the system realized this way tends to demand
very delicate handling for qubit operations
that often becomes quickly impractical in many-qubit settings.
In contrast, the quantum abacus model presented here possesses a semiclassical nature, 
as the system employs a qubit basis comprising infinitely many eigenstates, 
which is made possible by utilizing quantum caustics \cite{HM99,MT02}.  
As a result, the system can acquire robustness and spatial extendability,
which are highly desirable for quantum information processing 
beyond the level of laboratory experiments. 
The system we consider for qubit operations
consists of a particle of mass $m$ moving on a line ($-\infty < x < \infty$)
under a point interaction at $x = 0$ and a background harmonic oscillator potential with
frequency $\omega$ (FIG.~1).  
To describe a state $\psi(x)$ of the
system, it is convenient to introduce $\psi_+(x) = \psi(x)$ and
$\psi_-(x) = \psi(-x)$ for $x > 0$ and thereby
represent the wavefunction on the positive and
negative sides separately, using the two-component wavefunction
\begin{eqnarray}
\label{cj01}
\Psi(x) =
\begin{pmatrix}  \psi\plus(x)  \\  \psi\minus(x)  \end{pmatrix} ,
\qquad
x > 0.
\end{eqnarray}
The state of the particle is governed by the Hamiltonian
\begin{eqnarray}
\label{cj02}
H = -\frac{\hbar^2}{2m} \frac{\d^2}{\d x^2} + \frac{1}{2} m\omega^2 x^2
\end{eqnarray}
and is subject to the connection condition \cite{FT00,CF01,TF01}
\begin{eqnarray}
\label{cj03}
(U - I) \Psi(0) + i L_0 (U + I) {{\d\Psi} \over {\d x}} (0) = 0,
\end{eqnarray}
which is specified by the matrix $U \in U(2)$ characterizing the point interaction.
Here, $I$ is the two-by-two unit matrix, $L_0 \ne 0$ is a length
constant required on dimensional grounds, and
$\Psi$ and ${\d\Psi}/{\d x}$ are evaluated in the limit $x \to +0$.
%

%
\begin{figure}
\includegraphics[width=4.8cm]{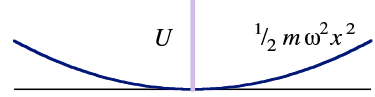}
\caption{\label{fig:epsart00}
Our setting for qubit operations: A harmonic oscillator potential 
with a tunable point interaction at the center
that provides a connection condition with a $U(2)$ matrix $U$.
}
\end{figure}
%
Among the point interactions described by (\ref{cj03}),
the familiar example of $\delta$-interaction,
which induces discontinuity in the derivative of the wavefunction,
is only a special
one-parameter subfamily.  In fact,
the $U(2)$ variety comprises such exotic point
interactions that cause discontinuity in the wavefunction itself, as
well as ones that admit a
constant transmission probability for all particle energies.
Explicit construction of these highly singular point interactions has
been achieved in terms of singular short-range limits of simple known
interactions \cite{CS98,EN01}.
To illustrate by some examples,
we mention that the identity
matrix $U = I$ results in $\frac{\d\psi\plus}{\d x\ } (0) $ $= \frac{\d\psi\minus}{\d x\ } (0)$ $ = 0$,
signifying a point interaction for an inpenetrable barrier at $x = 0$
with Neumann boundaries at its both sides.
Similarly, the negative
identity matrix $U = -I$ gives $\psi\plus (0) = \psi\minus (0) = 0$,
another inpenetrable barrier with Dirichlet boundaries.
Furthermore, it is the Pauli matrix $U = \sigma_1$ that provides
the free case ({\it i.e.}, no point interaction) with the smooth connection
conditions $\psi\plus(0) = \psi\minus(0)$, $\frac{\d\psi\plus}{\d x\ }(0) =
-\frac{\d\psi\minus}{\d x\ }(0)$.  Another important example is the case of
Hadamard matrix $U = (\sigma_1+\sigma_3)/\sqrt{2}$. It turns
out that this point interaction admits the constant transmission probability one-half
irrespective of the energy of the incident particle
\cite{CF01}.

A notable property of point interactions is that, given a state
$\Psi$ that fulfils (\ref{cj03}) with $U$, we obtain,
with an $SU(2)$ matrix $V$, 
a new state $\Psi' = V \Psi$  that satisfies (\ref{cj03}) with
\begin{eqnarray}
\label{cj04}
U' = V \, U \, V^{-1} .
\end{eqnarray}
The transformations by various $V$ provide a link between different 
point interactions, and lead to several salient properties among
the family of point interactions \cite{CF01,TF01}.
We utilize three of them, 
which are available when the background potential is parity invariant. 
%

%
The first is that if $\Psi$ is
an energy eigenfunction then $\Psi '$ is also an energy eigenfunction
with the same energy.
This implies that the two systems with their point interactions related by
the conjugation (\ref{cj04}) share the same energy
spectrum. Since those $U$ which fall into the same conjugacy class
under (\ref{cj04}) form a sphere
$S^2$, all such spheres in $U(2)$ are isospectral subfamilies.
The second property concerns the time evolution under point interactions $U$,
described by the transition operator
${\cal U}_t(U)=e^{- \frac{i}{\hbar} H t}$ as
\begin{eqnarray}
\label{cj05}
\Psi\,\, \timeevol{t} {\cal U}_t(U)\, \Psi.
\end{eqnarray}
Suppose ${\cal U}_t(U)$ is the transition operator under
the point interaction $U$, then states 
under $U'$ evolves with the transition operator
\begin{eqnarray}
\label{cj06}
{\cal U}_t(U') = V\, {\cal U}_t(U)\, V^{-1},
\end{eqnarray}
in correspondence to (\ref{cj04}).

The third is the fact that the conjugation (\ref{cj04}) 
can be used to link two physically distinct point interactions,
one separating (representing an inpenetrable barrier) and the other nonseparating,
relating thus a barrier totally reflecting from both
directions to a barrier which admits probability flow across $x = 0$.
Indeed, since any $U(2)$ matrix $U$ can be decomposed in the form 
\begin{eqnarray}
\label{cj07}
U = V^{-1}\, D\, V,
\end{eqnarray}
with $V \in SU(2)$ and a diagonal matrix $D$,
the conjugation (\ref{cj04}) with the same $V$ amounts to
the diagonalization of $U$ into $U' = D$.
If we parametrize $D$ as
\begin{eqnarray}
\label{cj08}
D
=
\begin{pmatrix}
e^{i \theta\plus} & 0 \\
0  & e^{i \theta\minus}
\end{pmatrix},
\end{eqnarray}
with $0 \le \theta\plusminus < 2 \pi$,
then the connection condition (\ref{cj03}) for $U = D$ reads
\begin{eqnarray}
\label{cj09}
\psi_\pm(0) + L_0\cot{\frac{\theta_\pm}{2}} \, {\d\psi_\pm\over{\d x}}(0) = 0,
\end{eqnarray}
showing that the positive and negative sides are separated physically.
Note that these $D$ in (\ref{cj08}) form  
a $U(1) \times U(1)$ torus subfamily in $U(2)$.
Since the spectrum is unchanged under conjugation, these $D$ index the
possible energy spectra of point interactions \cite{TF01}.
Visually, if the barrier represented by the point interaction is regarded as 
a gate, and if $U$ is non-diagonal allowing for transmission, then 
the conjugation $U \to D$ (or its inverse $D \to U$)
is equivalent to opening (or closing) the gate. 
In general, 
the conjugation $U \to U'$ by (\ref{cj04}) alters the transmission property of the gate 
while preserving the energy spectrum.
Hence, the conjugation provides a very useful means to control the gate, and plays
a vital role in the present proposal of quantum information processing.

%

Our realization of qubit operations employs
the class of point interactions given by
\begin{eqnarray}
\label{hh10}
U = V^{-1} \sigma_3 V ,  \qquad  V \in SU(2) .
\end{eqnarray}
This conjugacy class possesses no finite scale parameter
and provides (together with
$U = \pm I$) the subfamily of scale invariant point interactions in
the $U(2)$ family.  
Moreover, the class of 
the unitary matrices (\ref{hh10}) leads to the Bloch sphere when implemented by time evolution.  
This can be made more explicit by parametrizing
$V = e^{i \frac{1}{2} \mu \sigma_2} e^{i \frac{1}{2} \nu \sigma_3}$
with angles $0 \le\mu< \pi$ and $0 \le\nu< 2 \pi$ to obtain
$U = \sigma(c)$ where $\sigma(c)$ is a vector in the linear space 
spanned by the Pauli matrices,
\begin{eqnarray}
\label{hh11}
\sigma(c) = \sum_{i=1}^3
c_i\, \sigma_i
\end{eqnarray}
with 
$(c_1, c_2, c_3) = (\sin\mu\cos\nu, \sin\mu\sin\nu, \cos\mu)$.
In particular, if $\mu = \frac{\pi}{2},\, \nu = 0$
we have $U = \sigma_1$ which yields the smooth connection
condition.  The system then becomes the standard harmonic oscillator with
the energy eigenvalues
\begin{eqnarray}
\label{hh12}
\epsilon_n = (n + {1/2}) \hbar \omega,  \qquad  n = 0, 1, 2, \ldots
\end{eqnarray}
and the eigenfunctions
\begin{eqnarray}
\label{hh13}
\Phi_n^{\sigma_1}(x) =
\begin {pmatrix}  u_n(x)  \\  (-1)^n u_n(x)
\end{pmatrix} ,
\end{eqnarray}
where $u_n(x)$ are the familiar harmonic oscillator wavefunctions
consisting of Hermite polynomials.
Since the harmonic oscillator potential is parity invariant, all $U$
of (\ref{hh10}) share the same energy spectrum by the second 
property of conjugation mentioned before. 
The conjugation also
allows us to write the eigenstates $\Phi_n^{\sigma_3}$
for the point interaction $\sigma_3$ as
\begin{eqnarray}
\label{hh14}
\Phi_n^{\sigma_3} =
e^{i\frac{\pi}{4}\sigma_2} \Phi_n^{\sigma_1} =
\left\{ \begin{array}{cc}
\begin{pmatrix}  \sqrt{2} \, u_n  \\  0  \end{pmatrix}
\vphantom{\Bigg(}
& (n = {\rm even}) \\
\begin{pmatrix}  0  \\  - \sqrt{2} \, u_n  \end{pmatrix}
& (n = {\rm odd}) .
\end{array} \right.
\end{eqnarray}
We now consider the time evolution of the system under $U$ 
for a time step $\tau$, which we set to be 
the {\it half period of harmonic oscillation}, $\tau = {\pi} / \omega$.
{}For the case of point interaction $\sigma_3$, we find
\begin{eqnarray}
\label{hh15}
{\cal U}_\tau({\sigma_3}) \Phi_n^{\sigma_3} =
e^{- \frac{i}{\hbar} H \tau} \Phi_n^{\sigma_3} =
-i \, \sigma_3 \,\Phi_n^{\sigma_3} ,
\end{eqnarray}
for which the relation
$e^{- \frac{i}{\hbar} \epsilon_n \tau}$ $ =  -i (-1)^n$ is used.
Since an arbitrary state $\Psi$ is expressible as a linear combination of
$\Phi_n^{\sigma_3}$, its evolution under the point interaction $\sigma_3$
in half period $\tau$ is also given by ${\cal U}_\tau({\sigma_3})  \Psi$
$= -i \, \sigma_3 \Psi$.
Thus we obtain a remarkable expression
\begin{eqnarray}
\label{hh16}
{\cal U}_\tau({\sigma_3}) = -i\, \sigma_3,
\end{eqnarray}
namely, apart from the constant phase $-i$,
the time evolution operator under the point interaction $\sigma_3$ is
given by $\sigma_3$ itself.
With the conjugation (\ref{cj06}),
the time evolution of a state in a system under the scale invariant
family of point interaction $U$
is given by the evolution operator
\begin{eqnarray}
\label{hh17}
{\cal U}_\tau(U)
= - i U,
\end{eqnarray}
which is (up to $-i$) exactly
the matrix that characterizes the point interaction.
In brief, the spatial property encoded in a point interaction $U$
manifests itself in the time evolution ${\cal U}_\tau(U)$.
%
%
\begin{figure}
\includegraphics[width=7.7cm]{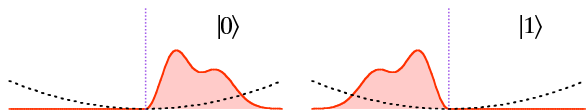}
\medskip
\caption{\label{fig:epsart02}
Our qubit basis of states $\left| 0 \right>$ and
$\left| 1 \right>$.  These are provided by a state of arbitrary profile 
localized on one side of the barrier and the other state given by the mirror image of the former.
}
\end{figure}
%

Note that
the present realization of unitary operations associated with the
Bloch sphere (\ref{hh10})
is available 
for an arbitrary state, not just for a particular pair of eigenstates.
This allows one to consider a qubit space spanned by a state 
with an {\it arbitrary} profile localized on one side of the barrier and its mirror state.    
A convenient choice for the 
qubit basis $\left| 0 \right>$ and $\left| 1 \right>$ is 
%
\begin{eqnarray}
\label{hh18}
\Psi_+ =
\begin {pmatrix}  f(x)  \\ 0
\end{pmatrix}
\leftrightarrow \left| 0 \right> ,
\quad
\Psi_- =
\begin {pmatrix}  0  \\ f(x)
\end{pmatrix}
\leftrightarrow \left| 1 \right> ,
\end{eqnarray}
where $f(x)$ is  an arbitrary real function 
having the normalization property $\int_0^\infty f(x)^2 \d x = 1$
(FIG. 2).
Since the spatial profiles of the
qubits  $\left| 0 \right>$ and $\left| 1 \right>$ are mirror symmetric,
any mixtures of $\left| 0 \right>$ and $\left| 1 \right>$ 
that emerge as a result of the time evolution ${\cal U}_\tau (U)$ have 
profiles that are analogous to each other
apart form the different scaling factors  
on both sides of the barrier (FIG. 3).  In other words, 
the state returns to the qubit space after the unitary 
evolution associated with the Bloch sphere for a time step $\tau$. 
%
\begin{figure}
\includegraphics[width=7.7cm]{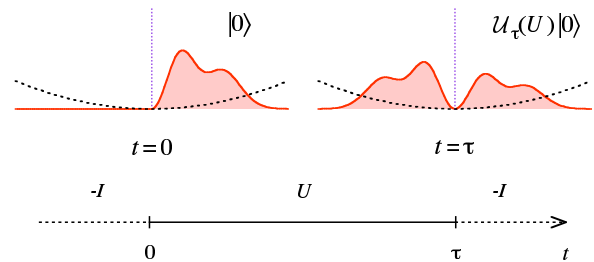}
\caption{\label{fig:epsart03}
Our qubit operation is achieved via the time evolution of the 
initial state that has an arbitrary profile.  For the time step $\tau$$=\pi/\omega$, 
the point interaction at the origin is 
switched from the closed gate $U = -I$ to an appropriate open gate $U$. 
The time evolution on the initial state, here illustrated by $\left| 0 \right>$, 
yields the unitary operation $U$ used for the opening gate by (\ref{hh17}).
The profile (modulo the scale) is kept after each time step.
}
\end{figure}
%

To implement an arbitary unitary operation beyond the Bloch sphere, 
we recall that a generic element  $U \in U(2)$ admits the decomposition (\ref{cj06}).  
It is easy to show that, instead of conjugation by $V$, a Bloch sphere element
$\sigma(c)$ with properly chosen $c_i$
may also be used to write 
$U = \sigma(c)\, D\, \sigma(c)$ \cite{TF01}.
The diagonal element $D$, in turn, can be expressed
as the product of two Bloch elements and an overall phase 
$D = e^{i\xi}\,\sigma(a)\,\sigma(b)$.  
This is achieved by using a set of
vectors $\sigma(a)$ and $\sigma(b)$ of the form (\ref{hh11}) 
which are orthogonal to each other in the plane perpendicular to the third axis,
as can be confirmed from the elementary identity
$\sigma(a)\,\sigma(b) 
= (a\cdot b) I + i \sum_i(a\times b)_i \sigma_i$.
Since the phase $e^{i\xi}$ may be cancelled out by adding a constant potential to 
the Hamiltonian (\ref{cj02}), 
we find that any unitary evolution $U$ can be performed essentially by
the successive application of corresponding time evolution operators belonging to 
the Bloch sphere in up to four steps, each of which is realized by the
time evolution for the half period $\tau$.  
We conclude, therefore, that 
by tuning the point interactions appropriately, all the $U(2)$ operations 
required for quantum processing can be implemented in the qubit space
spanned by $\Psi\plus$ and $\Psi\minus$.

There is a more direct method to implement 
the diagonal operation $D$, if we do not stick to the operation solely
by the point interactions.  
In order to obtain separate phases for $\left| 0 \right>$ and $\left| 1 \right>$, 
we can simply apply an extra constant potential 
on either side of the gate in addition to the overall constant potential,
while keeping the gate closed by setting $U=\pm I$.

Let us now consider the trasition between the classical 
and the quantam regime in our abacus operation.
The classical regime arises when 
we have a state of a sharply localized wave packet on which we consider 
only \lq classical\rq{} gates $U = - I$ and $U =  \sigma_1$ (up to the sign).
In each time step $\tau$, the wave packet, which can now be interpreted 
as a classical {\it bead}, bounces back at the gate 
and stays in the 
original side under the closed gate $U = - I$, or travels to the 
other side with the open (or NOT) gate $U= \sigma_1$.
Obviously, this is a microscopic realization of the classical abacus.

If, on the other hand, we bring all possible quantum gates $U \in U(2)$
in use, bilocal wave packets inevitably come into play, as seen in FIG. 3.
We then recover the full quantum abacus.
Thus, if we can implement the full set of quantum gates on a larger scale
--- which is possible as long as the harmonic potential 
is maintained accurately to the extent that
admits the caustic property --- we may
expect the quantum abacus to operate even in the semiclassical level.
The important point is that, when we read out the outcome
by the simple presence/absence of the particle in the
qubit carrier $\left| 0 \right>$ and $\left| 1 \right>$
(as in the case of final readout of Grover algorithm \cite{GR97}),
the precise profile of the qubit states does not matter so much.
Then, this robustness of the qubit operation 
will further be improved in the semiclassical regime.

The model considered here admits a further extension by the addition of
an extra repulsive $1/x^2$ potential, which keeps the structure of equi-spaced
energy levels intact, allowing for a quantum caustics to occur \cite{MT02}, and 
yet renders the wavefunctions localized away from the gate.
This will be useful for making the system more robust, especially in
the instantaneous switching process between an open and closed gates,
because the disturbance caused
by the switching is thought to occur in the area around the gate.

Once a qubit is realized, we can construct a multiple-qubit system by 
bringing many copies of our model system and string them together with 
a certain communication mechanism to allow the multiple-qubit operation.
The full treatment of this is outside the scope of the present work.
We simply mention one example of two-qubit Control-NOT operation 
realized by connecting two of our model systems by a trigger mechanism.  
Namely, if the particle in the first system is found in $x>0$ side, we set a signal 
to travel to the second system and
let the gate of the second system open ($U = \sigma_1$) for a time step 
$\tau$, while the gate is closed ($U=-I$) if no signal comes in.  The 
gate of the first qubit is kept closed during this operation.
%

In mathematical terms,
the two-qubit operations comprise a $U(4)$ group, and a more direct realization 
may be to manipulate the $U(4)$ family of point interactions that arise
in the form of \lq quantum $X$-junction\rq, a graph of four half lines
whose ends meet at a single point \cite{EX96}.

One obvious advantage of our implementation of
qubit operations is its visual and intuitive nature,
which brings a pedagogical value to our model. 
Another advantage is the robustness of the
qubit due to the simplicity of the setup, which is matched
only by the spin implementations.
In contrast, most solid-state based approaches
utilize particle states that tend to be
vulnerable to temperature fluctuations, whose suppression can
be costly and potentially inhibitory in constructing systems
with a larger number of qubits.
We hope that the model presented here offers 
a novel possibility for a robust and scalable
implementation of quantum computation.
\\

We thank Dr. M. Iwata and Dr. Y. Kikuchi
for helpful discussions.

%


\end{document}